%% file: ms.tex
\documentclass[runningheads]{llncs}
\pdfoutput=1
\newif\ifShowReviews
\ShowReviewsfalse

\RequirePackage[hyphens]{url} 
\usepackage{cmap} 
\usepackage{graphicx}
\usepackage{graphics}
\usepackage{authblk}
\usepackage{mathtools}
\usepackage{amsfonts}
\usepackage{subfigure}
\usepackage{amssymb}
\usepackage{acronym}
\usepackage{booktabs} 
\usepackage{tabularx} 

\usepackage{breakcites}
\usepackage[table,x11names]{xcolor}
\usepackage{amsmath,booktabs,mwe,pdflscape}
\usepackage{tablefootnote,footnotehyper}

\usepackage{xspace} 

\usepackage{hyperref}
\hypersetup{
    pdftitle={Influence Dynamics Among Narratives: A Case Study of the Venezuelan Presidential Crisis},
    pdfsubject={},
    pdfauthor={Akshay Aravamudan (akshay.aravamudan@gmail.com)},
    pdfkeywords={Influence Dynamics, Social Media Narratives, Multi-Variate Hawkes Processes, Granger Causality, Venezuelan Presidential Crisis},
    pdfcreator={LaTeX 2e},
    pdfproducer={pdfLaTeX},
    plainpages=false,			%
    unicode=false,				
    pdftoolbar=false,			
    pdfmenubar=true,			
    pdffitwindow=false,		
    pdfstartview={FitH},	
    pdfnewwindow=true,		
    urlcolor = blue 
}

\newacro{SM}{Social Media}
\newacro{TPP}{Temporal Point Process}
\newacroplural{TPP}{Temporal Point Processes}
\newacro{MVHP}{Multi-Variate Hawkes Process}
\newacro{PIM}{Process Influence Measure}
\newacro{SME}{Subject Matter Expert}
\newacro{LDA}{Latent Dirichlet Allocation}

\newcommand{\OURMEASURE}{\ac{PIM}\xspace}
\newcommand{\OURMEASURES}{\acp{PIM}\xspace}

\newcommand{\equref}[1]{Eq.~\ref{#1}} 
\newcommand{\secref}[1]{Sec.~\ref{#1}} 
\newcommand{\figref}[1]{Fig.~\ref{#1}} 
\newcommand{\tabref}[1]{Tab.~\ref{#1}} 

\newcommand\nnfootnote[2]{
    \stepcounter{footnote}
    \footnotetext{\url{#1}\label{fn:#2}}
}
\newcommand\blfootnote[1]{%
  \begingroup
  \renewcommand\thefootnote{}\footnote{#1}%
  \addtocounter{footnote}{-1}%
  \endgroup
}
\let\oldmaketitle\maketitle

\renewcommand{\maketitle}{\oldmaketitle\setcounter{footnote}{0}}

\begin{document}
\raggedbottom
    \input{TitleAuthors}
    \input{AbstractKeywords}

    \input{Introduction}
    \input{RelatedWorks}
    \input{Dataset}
    \input{Methodology}

    \input{Discussion}

    \input{Acknowledgments}

    \bibliographystyle{splncs04}
    \bibliography{references}

    \ifShowReviews
        \newpage
        \input{Reviews}
    \fi    

\end{document}

%% file: TitleAuthors.tex
\title{Influence Dynamics Among Narratives}
\subtitle{A Case Study of the Venezuelan Presidential Crisis}

\author{
Akshay Aravamudan \inst{1}
\and
Xi Zhang \inst{1}
\and
Jihye Song \inst{2}
\and
Stephen M. Fiore \inst{2}
\and
Georgios C. Anagnostopoulos \inst{1}
}

\institute{Department of Computer Engineering \& Sciences,\\Florida Institute of Technology, Melbourne, Florida, USA\\ 
\email{\{aaravamudan2014, zhang2012\}@my.fit.edu, georgio@fit.edu}\\ \and
University of Central Florida, Orlando, Florida, USA\\
\email{chsong@knights.ucf.edu, sfiore@ist.ucf.edu}}

\authorrunning{Aravamudan et al.}

\maketitle              

\blfootnote{Akshay Aravamudan, Xi Zhang, Jihye Song, Stephen M. Fiore, and Georgios C. Anagnostopoulos. Influence dynamics among narratives. In Robert Thomson, Muhammad Nihal Hussain, Christopher Dancy, and Aryn Pyke, editors, Social, Cultural, and Behavioral Modeling, 204–213. Cham, 2021. Springer International Publishing. doi: \href{https://doi.org/10.1007/978-3-030-80387-2_20}{\text{10.1007/978-3-030-80387-2\_20}}.}

%% file: AbstractKeywords.tex
\begin{abstract}

It is widely understood that diffusion of and simultaneous interactions between narratives --- defined here as persistent point-of-view messaging --- significantly contributes to the shaping of political discourse and public opinion. In this work, we propose a methodology based on Multi-Variate Hawkes Processes and our newly-introduced Process Influence Measures for quantifying and assessing how such narratives influence (Granger-cause) each other. Such an approach may aid social scientists enhance their understanding of socio-geopolitical phenomena as they manifest themselves and evolve in the realm of social media. In order to show its merits, we apply our methodology on Twitter narratives during the 2019 Venezuelan presidential crisis. Our analysis indicates a nuanced, evolving influence structure between 8 distinct narratives, part of which could be explained by landmark historical events.  

\keywords{Influence Dynamics \and Social Media Narratives \and Multi-Variate Hawkes Processes \and Granger Causality \and Venezuelan Presidential Crisis}

\end{abstract}

%% file: Introduction.tex
\section{Introduction}
\label{Introduction}

Discourse on social media platforms has become increasingly relevant in political contexts throughout the world \cite{Jungherr2016, Zhang2010}. There have been quite a few studies that have illustrated the power that social media yields in shaping public opinions and influencing political dialogues \cite{Jost2018, margetts2015}. While it has been argued that the Internet has not revolutionized how politics is conducted \cite{Hindman2009}, it does provide an efficient medium for rhetoric and discussion born out of existing cultural, economic, and political situations \cite{Calderaro2017}. As such, understanding the nature of how ideas, opinions, and calls to action diffuse over social media provides insights into how real-world events may unfold, especially in the context of areas experiencing political and economic turmoil. Such insights may help interested parties in developing a more comprehensive portrait of developing socio-geopolitical contentions.

The Venezuelan presidential crisis is an indicative example of significant political events. It began on January 10\textsuperscript{th}, 2019, when Nicol\'{a}s Maduro was sworn in for a second presidential term following a disputed election in May of 2018. This development prompted international reactions with nations siding with either Nicol\'{a}s Maduro or Juan Guaid\'{o}, then president of Venezuela's National Assembly. The situation escalated after Guaid\'{o} declared himself as the interim president on January 23\textsuperscript{rd}, 2019. All the while, there was significant participation on social media, which drove conversations globally. During this period, Twitter was heavily employed, resulting in competing opinions and, thus, rendering it as an effective tool for political communication \cite{Sytnik2020}.

We study this in the context of narratives associated with this event. We develop an interdisciplinary approach that unites social science theorizing on narratives with computational approaches to information evolution. Our narrative approach combines a parsimonious definition of online narratives (see \cite{Blackburn2020CorpusDF}) as ``\textit{... recurring statements that express a point of view}.'' We use this in conjunction with stances taken on a subset of narratives (\textit{i.e.}, support or disagree). In this way, we are able to model the mutual influence of complementary and competing narratives evolving over time. As such, we put forward an approach for quantifying and assessing the co-evolving influences among narratives under the prism of Granger causality as it is defined in the context of \acp{TPP}. This allows one to discern concrete influence motifs between online narratives. Such motifs serve as an additional facet of characterizing online discussions, which may prove useful to social scientists studying online discourse. We next describe work related to our approach. Then in \secref{sec:dataset}, we provide an overview of our dataset. 
\secref{sec:methodology} describes our methodology in some detail. As proof of concept, we then apply our methodology to study the influence dynamics of 8 concurrent narratives connected to the Venezuelan presidential crisis using Twitter data as outlined in \secref{sec:dataset}. Finally, we provide and comment on our findings in \secref{sec:discussion}. In particular, we are able to identify influence patterns that can be correlated to milestones of the crisis.


%% file: RelatedWorks.tex
\section{Related Works}
\label{sec:RelatedWorks}

Our work on narrative modelling falls under a broader umbrella of dynamic topic modelling for temporally sequenced documents, where topics correspond to narratives and documents to tweets. A first example of such an approach is temporal \acs{LDA} \cite{Wang2012}, which extends traditional \ac{LDA} and allows for learning topic transitions over time. A couple of other examples that we mention here are are based on \acp{TPP}, which are used to model topic dynamics. Lai et al. \cite{Lai2016} propose using a marked Hawkes process to uncover inter-topic relationships. Similarly, Mohler et al. \cite{Mohler2020} propose a Hawkes-Binomial topic model to detect mutual influence between online Twitter activity and real-world events. While our work also falls under this general strand, it focuses solely on learning topic dynamics (see \secref{sec:methodology}), since the topics themselves are ultimately provided by \acp{SME} as described in \secref{sec:dataset}. 


%% file: Dataset.tex
\section{Data}
\label{sec:dataset}

The Twitter dataset used for this study is a subset of a larger social media dataset curated by a data provider as part of a larger grant program and is described in \cite{Blackburn2020CorpusDF} and \cite{Blackburn2020Venezuela}. In accordance with the requirements of the organization funding the research, the data were anonymised prior to sharing it with researchers to protect user privacy. Overall, it encompasses over 7 million tweets, overwhelmingly in Spanish, from December 25\textsuperscript{th}, 2018, to February 1\textsuperscript{st}, 2019, a period which coincides with the commencement of the presidential crisis.

Each tweet has been annotated with potentially multiple narrative labels. The narratives were derived by applying topic modelling on the tweets' textual content via Non-negative Matrix Factorization combined with tf-idf statistics \cite{Xu2003}. These topics were subsequently refined and formalized by \acp{SME}. A subset of the tweets was then manually annotated by the same \acp{SME} and were then used to train a BERT-based multilingual cased multi-label classification model \cite{pires2019} that was then used to annotate the remaining tweets. A similar procedure was followed to label tweets as pro- or anti-Maduro. \footnote{Narrative and stance labelling was carried out by the data provider and was provided to us as is with very limited description of the process followed.}

\begin{figure}[htpb!]
    \centering
    \includegraphics[scale=0.18]{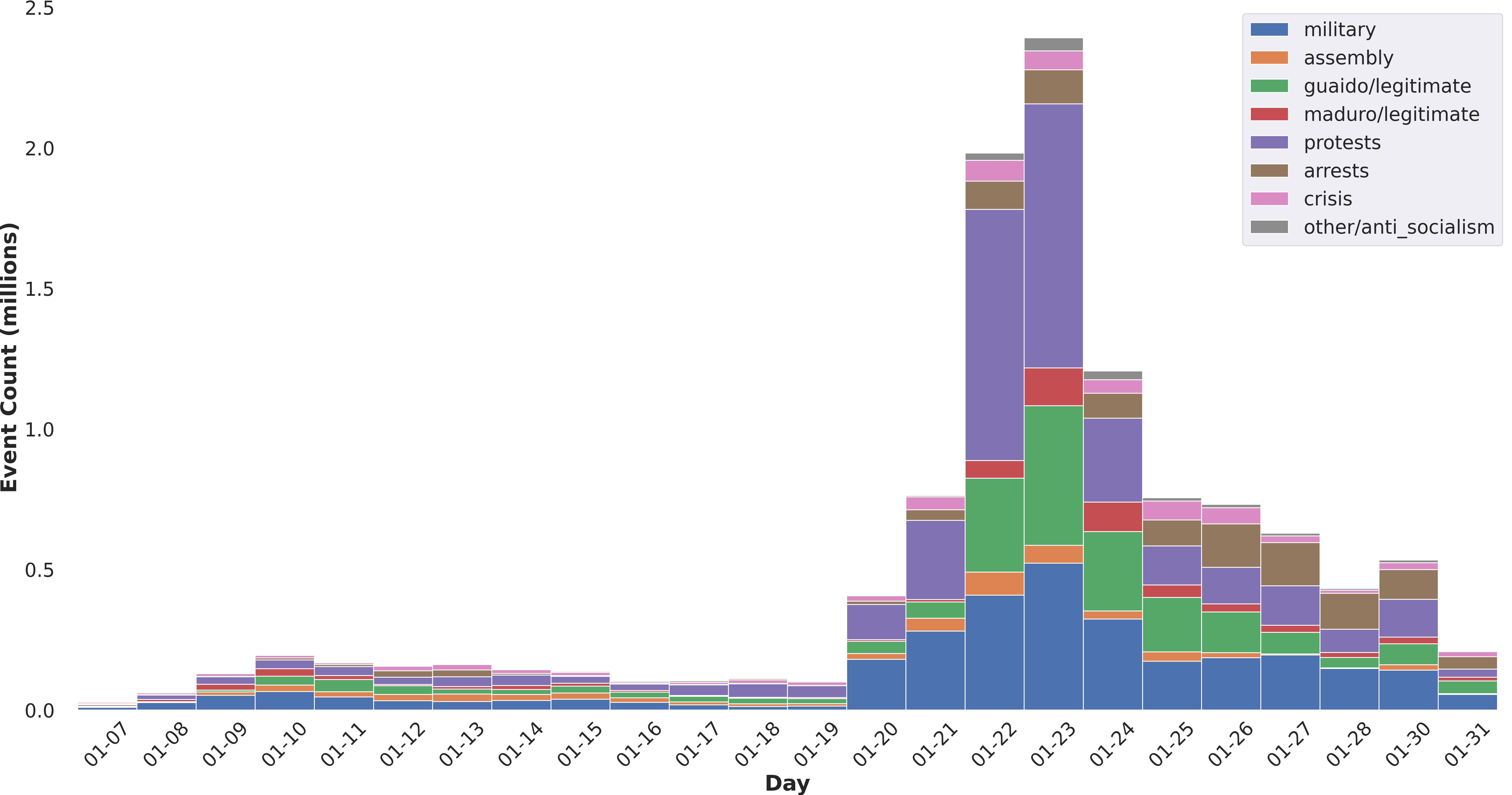}
    \caption{Histogram of Twitter event (tweet) counts per narrative in 2019. One observes a burst of activity between January 19\textsuperscript{th} and 21\textsuperscript{st}, during which there was a small-scale coup initiated by 27 soldiers.\textsuperscript{\ref{fn:1}} The peak activity occurring on January 23\textsuperscript{rd} appears strongly related to massive protests, which demanded Maduro to step down.\textsuperscript{\ref{fn:2}} During the same day, we also witness a significant increase in anti-Maduro tweets.}
    \label{fig:event_count_histograms}
\end{figure}

\addtocounter{footnote}{0} 
\nnfootnote{https://www.theguardian.com/world/2019/jan/21/venezuela-claims-foiled-attempted-military-uprising}{1}
\nnfootnote{https://www.theguardian.com/world/2019/jan/23/venezuela-protests-thousands-march-against-maduro-as-opposition-sees-chance-for-change}{2}

\renewcommand{\arraystretch}{1.2}
\begin{table}[htpb]
    \centering
     \caption{Stance distribution per narrative of the Venezuela Twitter data.}
    \begin{tabularx}{\textwidth}{@{} p{0.3\textwidth} p{0.233\textwidth} p{0.233\textwidth} p{0.233\textwidth} @{}}
        \toprule         
        \textbf{Narrative} & \textbf{Total Tweets} & \textbf{\% anti-Maduro} & \textbf{\% pro-Maduro} \\
        \midrule
        \addlinespace[5pt]
        \textit{military} & 1,534,242 & 67.38 & 21.87 \\
        \textit{assembly} & 252,448 & 95.79 & 1.82 \\
        \textit{guaido/legitimate} & 1,014,726 & 95.43 & 3.02 \\
        \textit{maduro/legitimate} & 304,127 & 2.92 & 96.72 \\
        \textit{protests} & 1,746,615 & 85.87 & 2.42 \\
        \textit{arrests} & 570,574 & 97.73 & 0.74 \\
        \textit{crisis} & 305,291 & 73.25 & 3.58 \\
        \textit{anti-socialism} & 101,716 & 78.04 & 14.77 \\
        \addlinespace[5pt]
        \bottomrule
        \addlinespace[5pt]
    \end{tabularx}
    \label{tab:data_description}
\end{table}

We only considered narratives that were present in at least 100,000 tweets. As a result, we analyzed a total of 8 narratives. The daily event counts of these narratives in the time period of interest is shown in \figref{fig:event_count_histograms}, while
\tabref{tab:data_description} shows the distribution of stances per narrative. The \textit{military} narrative includes discussions about the Venezuelan army, security services, or other organizations that reported to Maduro's government. \textit{Assembly} includes any mentions of Venezuela's National Assembly. \textit{Guaido/legitimate} and \textit{maduro/legitimate} consist of tweets that expressly support the legitimacy of Guaid\'{o} and Maduro, respectively. \textit{Protests} includes tweets that mention anti-Maduro demonstrations, public gatherings, or rallies. \textit{Arrests} includes tweets that refer to people who had been imprisoned at the time. Moreover, the \textit{crisis} narrative label refers to the Venezuelan humanitarian crisis\footnote{\url{https://www.reuters.com/article/us-venezuela-politics-un/venezuelans-facing-unprecedented-challenges-many-need-aid-internal-u-n-report-idUSKCN1R92AG}} and finally, \textit{anti-socialism} includes tweets that mention socialism, communism, or leftism as the primary cause of the humanitarian crisis.


%% file: Methodology.tex
\section{Methodology}
\label{sec:methodology}

\subsection{Multi-Variate Hawkes Processes}

In this work, we employed \acp{MVHP} \cite{Hawkes1971} --- a particular system of \acp{TPP} --- to characterize the temporal dynamics of tweets comprising the narratives of interest. \acp{MVHP} have been widely used for modelling social media events (\textit{e.g.}, \cite{Farajtabar2017, Zhou2013}) as they are capable of describing self- and mutually-exciting modes of event generation. \acp{TPP} are completely characterized by their conditional intensity $\lambda(t \mid \mathcal{H}_{t^-})$ at time $t$ given their past observations $\mathcal{H}_{t^-}$, which is referred to as the history of the process. The quantity $\lambda(t \mid \mathcal{H}_{t^-}) dt$ yields the expected number of events such a process will generate in the interval $(t, t+dt)$. Assume an \ac{MVHP} comprising $P$ processes. Its $i$\textsuperscript{th} process features a conditional intensity of the form
\begin{align}
    & \lambda_i (t \mid \mathcal{H}_{t^-}) =  b_i(t) + a_{i,i} \sum_{\mathclap{t_k^i \in \mathcal{H}_{t^-}^i}}  \phi_{i,i}(t - t_k^i) + \sum_{\substack{j \in \mathcal{P}\\ j \neq i}} \alpha_{i,j}  \sum_{\mathclap{t_k^j \in \mathcal{H}_{t^-}^j}}  \phi_{i,j}(t - t_k^j)
    \label{eq:conditional_intensity}
 \end{align}
where $i \in \mathcal{P} \triangleq \{ 1, 2, \ldots, P \}$, $\mathcal{H}_{t^-}^i$ is the history of the i\textsuperscript{th} process and $\mathcal{H}_{t^-} \triangleq \cup_{i \in \mathcal{P}} \mathcal{H}_{t^-}^i$ is the history of the multi-variate process. In \equref{eq:conditional_intensity},  $b_i(\cdot) \geq 0$ is a history-independent intensity component, which we will refer to as base or background intensity. The quantity $\phi_{i,j}(t - t_k) \geq 0$ reflects the (typically, momentary and, subsequently, diminishing) increase in intensity that the $i$\textsuperscript{th} process will experience at time $t$ due to an event of the $j$\textsuperscript{th} process, which occurred at time $t_k$. The functions $\phi_{i,j}(\cdot)$ are referred to as memory kernels. Also, for fixed base intensity and memory kernels, the non-negative $a_{i,j}$'s constitute the \ac{MVHP}'s model parameters. Finally, apart from the background intensity, one can distinguish two additional terms in \equref{eq:conditional_intensity}: \textbf{(i)} a second, self-exciting term, through which past events of the process may generate further future events and \textbf{(ii)} a third, cross-exciting term, through which events of the other processes may do the same. An \ac{MVHP}'s base intensity and memory kernel functions are typically inferred non-parametrically, while the $a_{i,j}$ parameters are learned via (sometimes, penalized) maximum likelihood estimation. 

For the purposes of studying influences among narratives, we used \acp{MVHP} that featured one process per narrative. Each process utilised a constant background intensity $b_i$ and process-independent memory kernels $\phi_{i,j}(t) = e^{-t}$, which is a typical choice for Hawkes processes. Moreover, we opted to fit such \acp{MVHP} using overlapping time frames within the period of interest for two reasons: \textbf{(i)} initial experimentation gave strong indications---via probability-probability plots---that fitting a single \ac{MVHP} to the entire period would result in a bad-fitting model. It appears that the tweet dynamics under consideration are not well approximated by \acp{MVHP} at long time ranges, but only at shorter ones. And, \textbf{(ii)} modeling tweet dynamics on overlapping (instead of disjoint) time frames has a beneficial smoothing effect on the estimated model parameters, when comparing time-adjacent models. 

\subsection{Quantifying \& Comparing Influence Effects}
\label{sec:QuantifyingComparingInfluenceEffects}

Based on the notion of causality for continuous stochastic processes, Eichler et al. \cite{Eichler2017} extend the definition of Granger (predictive) causality to \acp{MVHP} and show that, if $a_{i,j} > 0$, then the $j$\textsuperscript{th} process Granger-causes the $i$\textsuperscript{th} process. This allows one to discern influences within and between processes by mere inspection or, more formally, by testing against $a_{i,j} = 0$. Moreover, the base intensity of each process can be naturally interpreted as a nonspecific cause, which is external to the system of processes. One can reject the absence of such unaccounted-for cause by testing against $b_i = 0$. Of course, the entirety of this causal reasoning is predicated on the absence of other confounding processes (or factors, in general), which is an assumption that we will also adopt for our studies. Thence, in our exposition we will make use of the notion of apparent influences among processes as a stand-in for Granger-causal effects sans confounding.

While testing for $a_{i,j} = 0$ determines a potential influence via a binary decision, quantifying the magnitude of the influence effect based on the $a_{i,j}$ parameters in order to draw comparisons is a much more nuanced issue. Some prior work, such as \cite{Alvari2019} and \cite{Lai2016}, for example, attempt to directly interpret the parameters' magnitudes as corresponding magnitudes of influence for the purpose of comparisons. This approach, though, is fraught with problems: \textbf{(i)} the first one is semantic in nature: the physical meaning of the $a_{i,j}$'s is intrinsically linked to the specific form of the memory kernels they multiply and, hence, is often far from straightforward to describe. And, \textbf{(ii)} solely relying on the $a_{i,j}$'s values to quantify the influence effect completely ignores memory kernels, which are equal contributors in shaping event dynamics; based on this approach, this renders comparing influence effects problematic.     

In order to circumvent these shortcomings, we introduce and advocate the use of \OURMEASURES for quantifying the magnitude of process-to-process influences. In particular, we define the $j$-to-$i$ \OURMEASURE $\pi_{i,j} \in [0,1]$ as the probability that an event of the $j$\textsuperscript{th} process is the most likely cause for an event of the $i$\textsuperscript{th} process. These probabilities can be easily estimated from the results provided by a fitted model. This is due to the following fact that readily stems from the theory of \acp{TPP}, whose conditional intensity is represented additively: if $E_k^i$ is the $k$\textsuperscript{th} event of process $i$, which occurred at time $t_k^i$, and $\mathcal{E}^j$ denotes the set of all events of process $j$, then
\begin{align}
    & \mathbb{P} \left\{ \text{$E_k^i$ was caused by any earlier event in $\mathcal{E}^j$} \right\} = \frac{ a_{i,j} \sum\limits_{ \mathclap{ t_{\ell}^j \in \mathcal{H}_{{t_k^i}^-}^j } } \phi_{i,j}(t_k^i - t_{\ell}^j)  }{\lambda_i(t_k^i \mid \mathcal{H}_{{t_k^i}^-}^i ) }
    \label{eq:trigerring_source_expression}
\end{align}
\begin{align}
    & \mathbb{P} \left\{ \text{$E_k^i$ was caused by $b_i(t)$} \right\} = \frac{b_i(t_k^i)}{\lambda_i(t_k^i \mid \mathcal{H}_{{t_k^i}^-}^i)}
    \label{eq:trigerring_background_source_expression}
\end{align}

for $i, j \in \mathcal{P}$. It is important to note that these probabilities jointly depend on model parameters, memory kernels as well on relative event timings and, thus, capture multiple facets of process dynamics. One can iterate over all events of process $i$ and, each time, record the process $j$, which exhibits the largest probability given by \equref{eq:trigerring_source_expression}. Then, the \OURMEASURE $\pi_{i,j}$ is estimated via the frequency with which an event of process $j$ was the likeliest cause for an event of process $i$.


%% file: Discussion.tex
\section{Results \& Discussion}
\label{sec:discussion}

\begin{figure}
\centering     
\subfigure[12--13 January 2019]{\label{fig:a}\includegraphics[width=60mm]{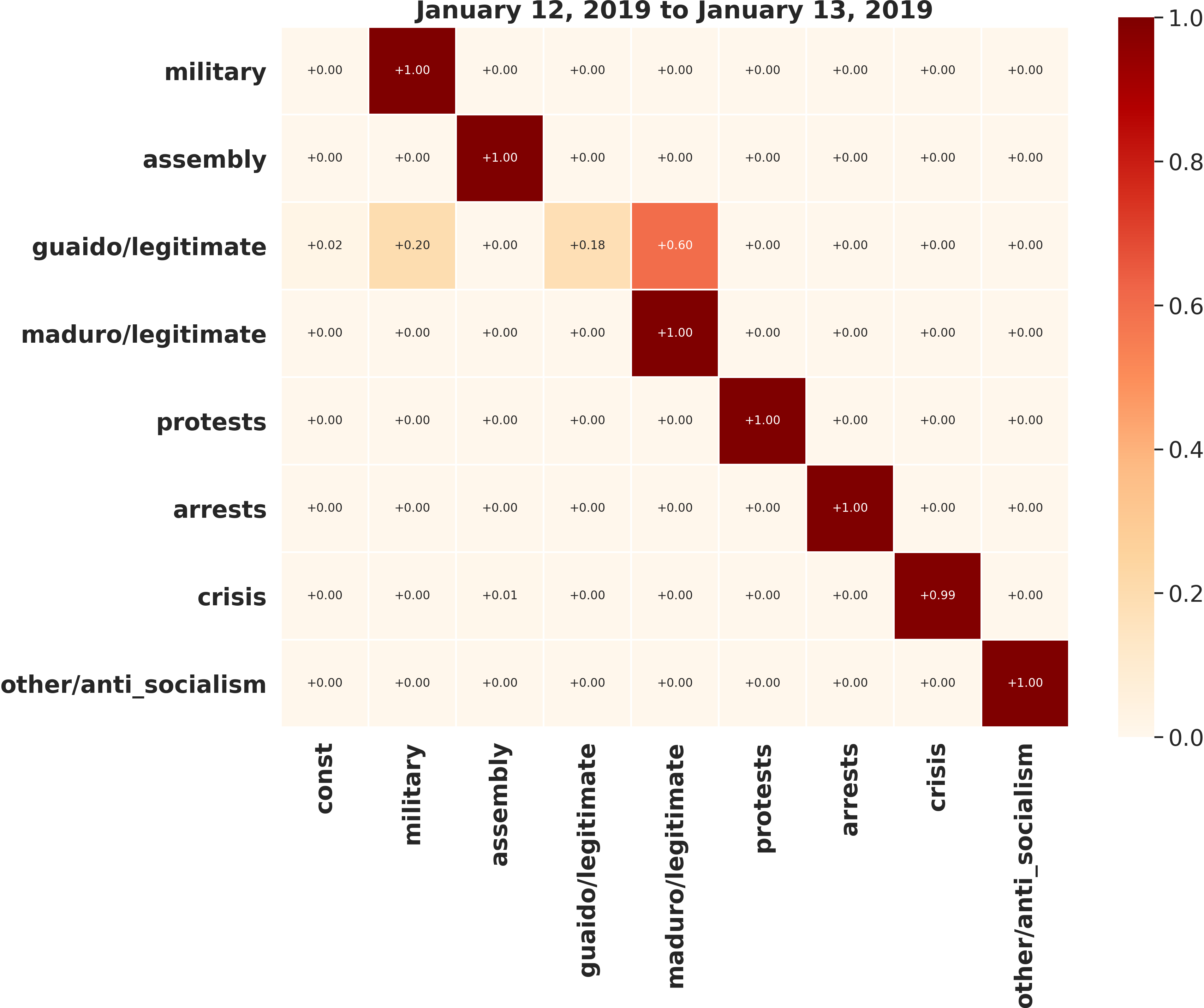}}
\subfigure[ 22--23 January 2019]{\label{fig:b}\includegraphics[width=60mm]{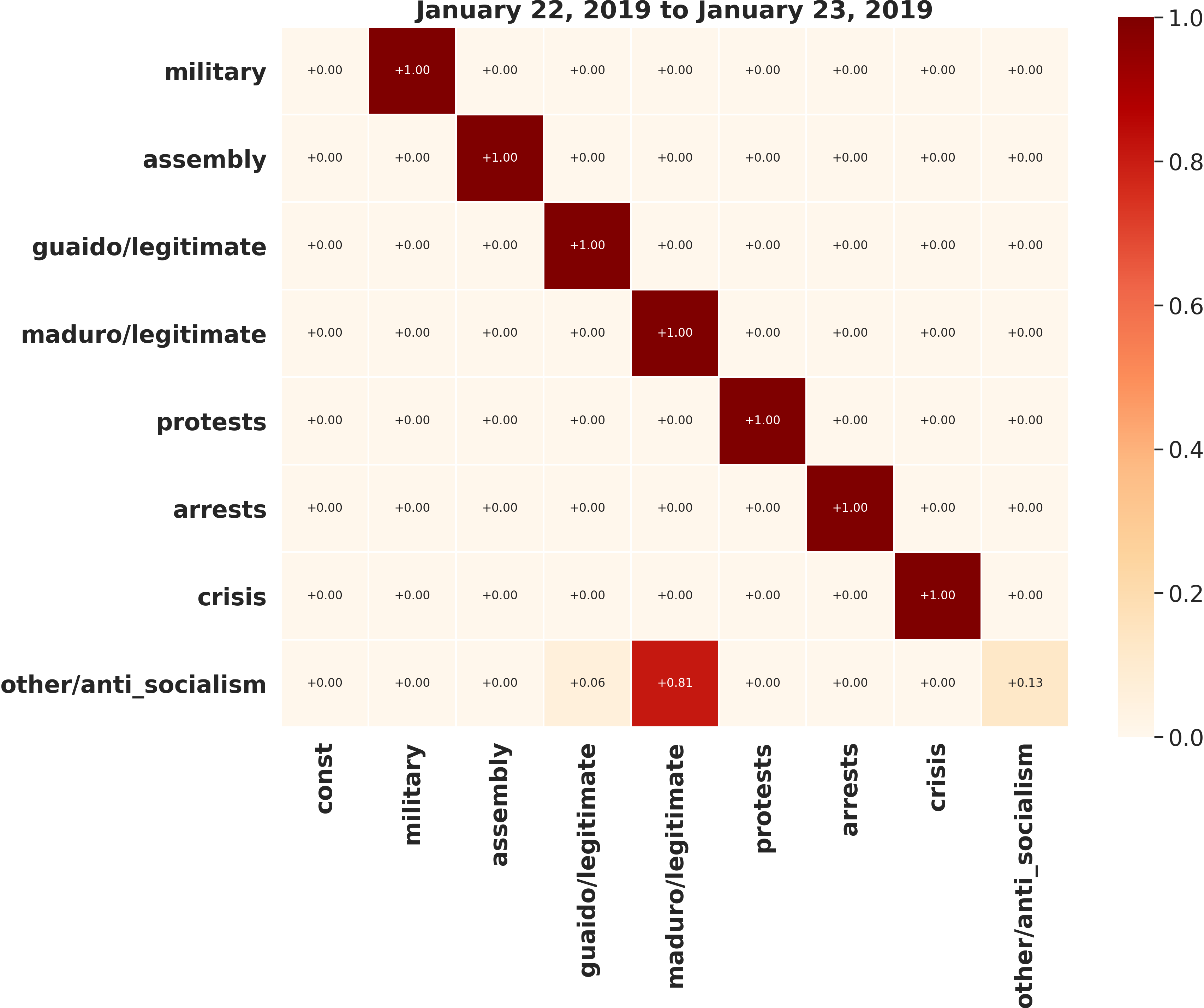}}
\caption{\OURMEASURE heat maps for two time frames. Columns show influence sources, while rows depict the narratives we studied. These maps illustrate self-driving narratives (prominent diagonal entries), as well as inter-narrative influences (sizeable off-diagonal entries).}
\label{fig:probability_heatmaps}
\end{figure}

\begin{figure}[htpb]
    \centering
    \includegraphics[width=\textwidth]{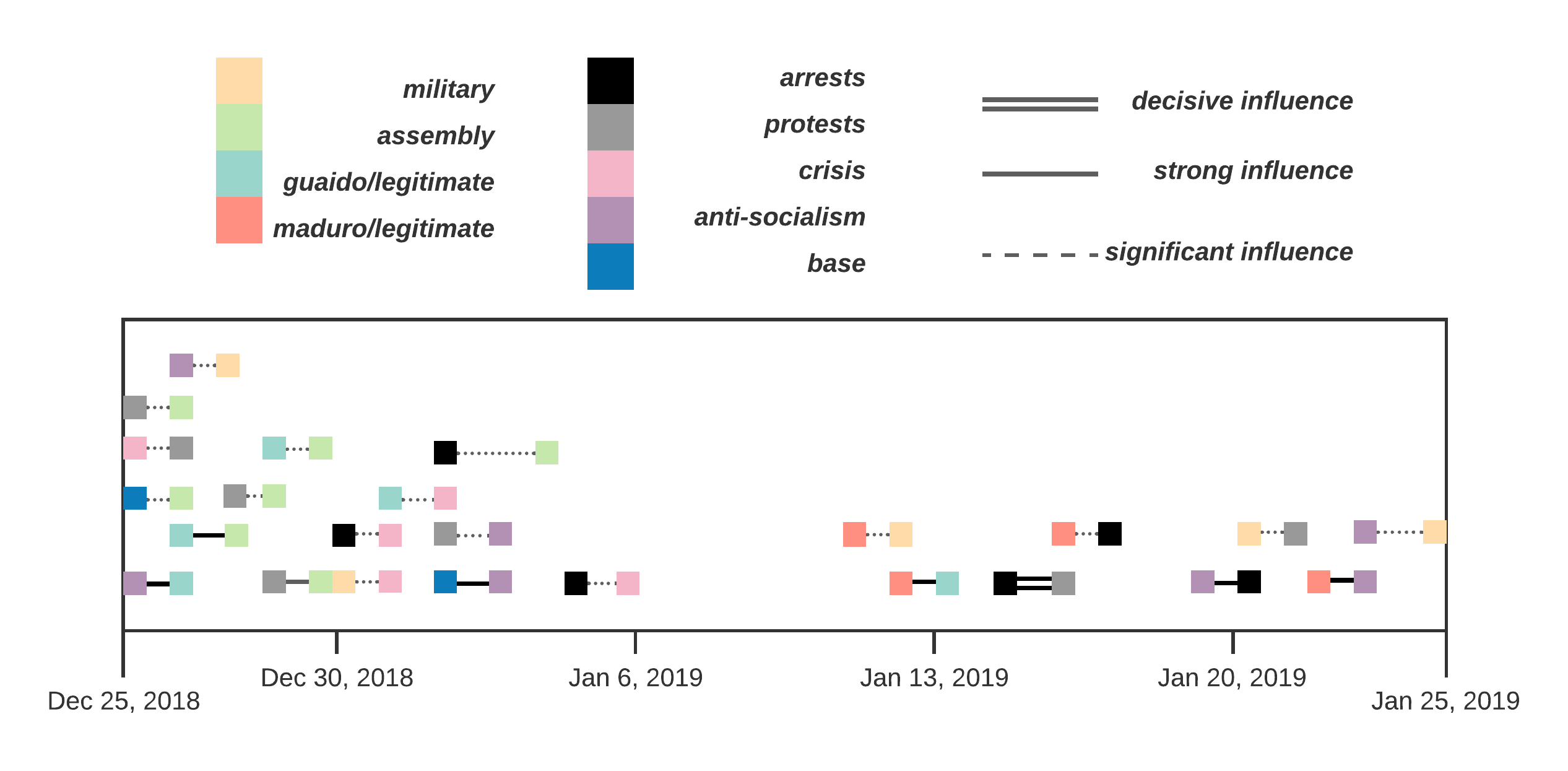}
    \caption{Timeline indicating noteworthy influences based on our estimated \OURMEASURES. Weaker cross-narrative influences with \OURMEASURE values $\pi_{i,j} \in [0.0, 0.2]$ and self-influences of any kind have been omitted for clarity. Note that, during the period of January 6\textsuperscript{th} through the 11\textsuperscript{th}, narratives only influence themselves ($\pi_{i,i} \approx 1$).}
    \label{fig:timeline_influences}
\end{figure}

We used the setup described in \secref{sec:methodology} to infer intra/inter-narrative influences from the available data. We split the observed time period into 35 two-day time frames that were overlapping by one day.\footnote{ In particular, we chose two-day time frames to reduce the computational burden of training. Also, we used an hourly timescale to represent event time stamps to maintain the numerical stability of our training algorithm.} For each such time frame, we trained a \ac{MVHP} and ensured that it was exhibiting (at least) a reasonably good fit as judged by probability-probability plots. For investigating influences among narratives, each \ac{MVHP} encompassed a process per narrative. Furthermore, we computed and tabulated \OURMEASURES as described in \secref{sec:QuantifyingComparingInfluenceEffects}. \figref{fig:probability_heatmaps} provides \OURMEASURES as heat maps for two indicative time frames: one before Guaid\'{o}'s  self-proclamation as legitimate president and one after. These particular examples illustrate self-reinforcing narratives ($\pi_{i,i} \approx 1$ diagonal values), as well as narratives being influenced by one or more of the remaining narratives (appreciable $\pi_{i,j}$ off-diagonal values). We characterize influences in terms of \OURMEASURES as follows: ``significant'' influences, when $\pi_{i,j} \in (0.2, 0.6]$ , ``strong'' influences, when $\pi_{i,j} \in (0.6, 0.99]$ and ``decisive,'' when $\pi_{i,j} > 0.99$. \figref{fig:timeline_influences} highlights such influences over the time frame that our study considered.

Judging from the results in \figref{fig:timeline_influences}, we notice that \textit{anti-socialism} strongly influences \textit{guaido/legitimate} between December 25\textsuperscript{th} and 26\textsuperscript{th}, which may reflect Guaid\'{o}'s support from Western governments and the growing dissatisfaction of the Venezuelan populace with existing conditions under Maduro's government, which were prevalent even before the crisis. Moreover, after Maduro's inauguration on January 10\textsuperscript{th}, we notice that \textit{maduro/legitimate} significantly influences \textit{military}. The tweets during these days were posted after the first open cabildo --- a political action convention --- held by then president of the National Assembly, Guaid\'{o}, where he was voted in as acting president. This could have motivated Maduro supporters to favor a strengthening of his claim to the presidency and, by extension, favor the military's response to the  unrest.\footnote{\url{https://www.bbc.com/news/world-latin-america-47036129}} Between January 12\textsuperscript{th} and 13\textsuperscript{th}, \textit{maduro/legitimate} significantly affects \textit{guaido/legitimate} showing a reactionary response to the mobilisation of Maduro's military in response to rising support for Guaid\'{o}. This also coincides with Guaid\'{o}'s brief arrest\footnote{\url{https://www.nytimes.com/2019/01/13/world/americas/venezeula-juan-guaido-arrest.html}} by the Bolivarian Intelligence Service on January 13\textsuperscript{th}. Finally, we notice that between January 20\textsuperscript{th} and 21\textsuperscript{st}, \textit{military} significantly influences \textit{protests}, which overlaps with when some National Guardsmen rose against Maduro.\footnote{\url{https://www.nytimes.com/2019/01/21/world/americas/venezuela-maduro-national-guard.html}} This was followed by a widespread and vociferous protests against Maduro. Between January 24\textsuperscript{th} and 25\textsuperscript{th}, we observe a significant influence from \textit{anti-socialism} to \textit{military}, which may be attributed to the wake of anti-Maduro protests\footnote{\url{https://www.theguardian.com/world/2019/jan/23/venezuela-protests-thousands-march-against-maduro-as-opposition-sees-chance-for-change}} on January 23\textsuperscript{rd} that called for the military to relinquish their allegiance to Maduro.

Finally, a potentially interesting extension of our work could specifically address strategic platform manipulation that may drive narratives during political crises. For instance, an operation involving Facebook and Instagram accounts attributed to a U.S. communications firm was flagged and removed for coordinated inauthentic behavior targeting Venezuela\footnote{\url{https://about.fb.com/news/2020/09/august-2020-cib-report/}} and was found to have primarily posted anti-Maduro content \cite{cryst_bolivarian_2020}. Additionally, Twitter announced the removal of a large Venezuelan state-backed operation from its platform.\footnote{\url{https://blog.twitter.com/en_us/topics/company/2019/further_research_information_operations.html}} Given the role of the Venezuelan political crisis in international geopolitics, future work could examine how domestic and foreign actors are leveraging social media to manipulate narratives for political objectives. To achieve this, we could use a more elaborate \ac{MVHP} to model such actors as separate processes. 

In summary, we have empirically demonstrated via our Venezuelan case study that \OURMEASURES, as introduced in \secref{sec:QuantifyingComparingInfluenceEffects}, furnish an unambiguous and interpretable dynamic characterisation of intra-/inter-narrative influences. Furthermore, we have showcased how such dynamics may be explained by landmark events --- exogenous to social media --- within broader historical contexts. In this capacity, our work provides an additional, important lens for studying influence between narratives, which, we hope, may prove useful to social scientists.


%% file: Acknowledgments.tex
\section*{Acknowledgments}

This work was supported by the U.S. Defense Advanced Research Projects Agency (DARPA) Grant No. FA8650-18-C-7823 under the \textit{Computational Simulation of Online Social Behavior} (\textit{SocialSim}) program of DARPA's Information Innovation Office. Any opinions, findings, conclusions, or recommendations contained herein are those of the authors and do not necessarily represent the official policies or endorsements, either expressed or implied, of DARPA, or the U.S. Government. Finally, the authors would like to thank the manuscript's anonymous reviewers for their helpful comments and suggestions.

%% file: Reviews.tex
\section*{Reviews}
\label{sec:Reviews}

\subsection*{Review \#1}

\textbf{Score:} {\color{Green4} 3 (strong accept)}

\medskip

\noindent
\textbf{Comments:} \\
The paper is a strong accept for the following reasons: shows empirical evidence and significant findings of testing out newer methods in dynamic topic analysis.
\noindent
\textbf{Response:}\\
No response. 
\subsection*{Review \#2}

\textbf{Score:} {\color{Green2} 2 (accept)}

\medskip

\noindent
\textbf{Comments:} \\
The methodology proposed in this article is quite eye-catching. The application of Multi-Variate Hawkes Processes provides along with the Process Influence Measures provides potentials for following researchers to analyze the interactions between topics in large corpora. However, I think a big difficulty is the first step: finding out consistent and informative topics from the corpora. The authors used manually annotated topics in this research, which might not be applicable in more general scenarios, especially when the scope of potential topics is not pre-refined. Another point I am doubt about is the choice of the time unit. As the author suggested that the time frame for fitting the model is 2-days, and a larger data frame would have terrible fits, I hope the authors could talk more about their experience in choosing the time frame and the time unit of the tweets. The final question is that the method is used with the assumption that the change of intensity of topics is purely driven by base intensity and inter/intra-topic interactions. But as the authors found evidence of real events to map their findings, I suspect these real events are also important factors that directly influence the intensity of topics. So there might be more explanations about the identification or distinction between the causality between the real events and topics/discourse and the causality purely from topic interactions.

\noindent
\textbf{Response:}\\
Regarding the first item about extracting the data in general scenarios, we agree and this, in our opinion, is very difficult and untrustworthy to extract without manually annotating the data. Moreover, since we do not have control over this process, refining it further is not practical.

Regarding the second item about the choice of time unit, we have incorporated the time units used. We used the 2-day time frame due to computational considerations. This detail has also been incorporated into the manuscript. 

In regards to the final point, the fact that real world events influence the events themselves is a limitation of this study. As a matter of fact, these events could be themselves incorporated into our setup to discern their influences to relevant narratives. This will be considered in an expanded version of our work.

\subsection*{Review \#3}

\textbf{Score:} {\color{Gold1} -1 (weak reject)}

\medskip

\noindent
\textbf{Comments:} \\
This work is well-written and proposes a new “Process Influence Measures” for quantifying and assessing how multiple socio-geopolitical narratives interact and influence each other. As the authors rightly point out, such an analysis would be very powerful and could lead to strong results. However, such strong results need to be well grounded with suitable evidence that need to be brought out by the analysis. Though this work uses Granger-causality model to identify these narrative evolutions, it does not provide sufficient analysis and evidence to justify the conclusions. Below are a few references to text and concerns.

\begin{enumerate}

    \item ``\textit{Our narrative approach combines a parsimonious definition of online narratives (see [2]) as ``... recurring statements that express a point of view.'' We use this in con- junction with stances taken on a subset of narratives (i.e., support or disagree). In this way, we are able to model the mutual influence of complementary and competing narratives evolving over time.}''
    
    \smallskip
    
    This is an interesting and strong definition for narrative. Some example tweets that belong and don’t belong to a narrative would be good. 
    
    \smallskip
    
    \item ``\textit{A subset of the tweets was then manually annotated by the same SMEs and were then used to train a BERT-based multilingual cased multi-label classification model [15] that was then used to annotate the remaining tweets. A similar procedure was followed to label tweets as pro- or anti-Maduro.}''
    
    \smallskip
    
    How good is the classifier? No evaluation/analysis provided.

    \smallskip

    \item  ``We only considered narratives that were present in at least 100,000 tweets. As a result, we analyzed a total of 8 narratives.''
    
    \smallskip
    
    Did tweets belong to more than 1 topic? How were they treated?
    
    \smallskip
    
    \item ``\textit{The military narrative includes discussions about the Venezuelan army, security services, or other organiza- tions that reported to Maduro’s government. Assembly includes any mentions of Venezuela’s National Assembly. Guaido/legitimate and maduro/legitimate con- sist of tweets that expressly support the legitimacy of Guaido and Maduro, respectively. Protests includes tweets that mention anti-Maduro demonstrations, public gatherings, or rallies. Arrests includes tweets that refer to people who had been imprisoned at the time. Moreover, the crisis narrative label refers to the Venezuelan humanitarian crisis3 and finally, anti-socialism includes tweets that mention socialism, communism, or leftism as the primary cause of the humanitarian crisis.}''
    
    \smallskip
    
    Was it a simple keyword based filtering to group tweets into narratives? That might be too weak to model a complex narrative described above.
    
    \smallskip
    
    \item More analysis is required to understand if the MVHP can identify evolutions of complex social media narratives and if so, what do the results imply.

\end{enumerate}

\noindent
\textbf{Response:}\\
The response to each sub-item is provided below.
\begin{itemize}
    \item We are unable to incorporate examples in this document, but we intend to do so in an extended version of this manuscript.
    \item We don't have this information currently since the data procurement data labelling was done by a third party data provider. We will get more information on this and include it in an extended version of the manuscript. 
    \item Yes, tweets could belong to more than one topic. In such cases, we treated the tweets separately per narrative. This was done since our intention was to obtain influences between narratives and would defeat the purpose of our intent if not treated separately. This has been covered in the data section.
    \item The method of keyword extraction utilises a combination of LDA with input from Subject Matter Experts.
    \item We agree that more analysis is needed to juxtapose the identified influences by virtue of analysing the text that contributed to this influence. This will be detailed in an extended version of the manuscript.
\end{itemize}

\subsection*{Review \#4}

\textbf{Score:} {\color{Green2} 2 (accept)}

\medskip

\noindent
\textbf{Comments:} \\
Overall, I think the approach is very interesting and likely useful as a way of understanding narratives derived from Twitter and other social media data. The only reason I did not rate it a Strong Accept is because I had trouble understanding the process used for analyzing the data, and I'm not sure I could replicate it given solely the information presented in the paper. While that may certainly be simply my own limitation, I think it would have been helpful to do a clearer walkthrough of the process. However, a conference presentation is an excellent place to do that, and I believe this paper both describes its approach well enough and prevents a novel enough method that it's worthwhile accepting it for publication.

\medskip

There is one thing I would recommend providing more clarity on before publication, if possible. Unless I missed it, there isn't a description of what languages the tweets were in, nor where they were from geographically. I assume that is described in detail in the papers referenced as the source of the data, but those are important points to at least mention in this paper.

\noindent
\textbf{Response:}\\
We will better explain the process in the presentation as well as an extended version of this manuscript. We will also detail the process to be able to reproduce these results. 

Regarding the final point, we observed that Spanish was the dominant language, a point we have included in the data section.

\subsection*{Tasks for an extended version of the paper}

Given the above reviews and taking into consideration the 10 page limit

\begin{itemize}
    \item [-] Talk about experiences in choosing time frame and the chosen time unit of processing tweets (hours)
    ``We chose the two day time frame after taking into consideration both the volume of tweets (see figure 1) in consideration and the running time of the algorithm''
    \item [-] Talk about possibility of external events influencing the tweets and how that is realized in our analysis of influences between tweets. More analysis to identify evolutions of complex social media narratives. What do the results imply ?
    ``Since external landmark events haven't been explicitly modelled, they can adversely affect each of the narratives and would manifest itself as narratives influencing other narratives. This, however, sets the stage for incorporating external events and determining their influence to these narratives. ''
    \item [2] Include examples of tweets belonging to narratives.
    
    \item [-] Quality of the BERT classifier for the multi-label classification of the tweets.
    \item [-] More detailed description of the how narratives are chosen for a single tweet.
    \item [-] Explain more clearly how one can replicate this method for analyzing data.[a clearer walk through of the process]
    \item [-] Explain more clearly the languages that prevail in the dataset.

\end{itemize}